\documentclass[10pt]{article}

\usepackage[body={16cm,23cm}]{geometry}
\usepackage{epsfig,amsfonts,amssymb}
\usepackage{enumerate}
\usepackage{cite}

\def\Dcal{\mathcal{D}}

\def\a{\alpha}
\def\b{\beta}
\def\g{\gamma}
\def\d{\delta}

\def\t{\theta}
\def\vt{\vartheta}
\def\beq{\begin{equation}}
\def\eeq{\end{equation}}
\def\be{\begin{displaymath}}
\def\ee{\end{displaymath}}
\def\bea{\begin{eqnarray}}
\def\eea{\end{eqnarray}}
\def\ov{\overline}
\def\bmat{\left(\begin{array}}

\def\ds{\displaystyle}

\def\Wp{ \raise.4ex\hbox{\textrm{\Large $\wp$}}}
\def\QQ{\mathcal Q}
\def\f{\Psi}

\def\P{{\mathcal P}}
\def\q{{\mathsf q}}

\def\eqref#1{(\ref{#1})}
\def\?{(?)\marginpar{|?}}

\newtheorem{conj}{Conjecture}
\begin{document}

\title{The eight-vertex model and Painlev\'e VI}

\author{        V V Bazhanov\footnote{email:
                {\tt Vladimir.Bazhanov@anu.edu.au}},
                V V Mangazeev\footnote{email:
                {\tt vladimir@maths.anu.edu.au}}\\ \\
 Department of Theoretical Physics,\\
         Research School of Physical Sciences and Engineering,\\
    Australian National University, Canberra, ACT 0200, Australia.}

\maketitle

\begin{abstract}
In this letter we establish a connection of Picard-type
elliptic solutions of Painlev\'e VI equation with the special
solutions of the non-stationary Lam\'e equation. The latter appeared in the
study of the ground state properties of Baxter's solvable
eight-vertex lattice model
at a particular point, $\eta={\pi}/{3}$, \ of
the disordered regime.
\end{abstract}

\section{Introduction}\label{eight}
The Painlev\'e transcendents have numerous remarkable applications in the
theory of integrable models of statistical mechanics and quantum field
theory (see, for instance, \cite{BMW,JMSM,CV91}).
Mention, in particular, the calculation of the ``supersymmetric
index'' and related problems of dilute polymers on a cylinder
which lead to Painlev\'e III \cite{CF92,Zam94}.
These problems are connected with
the finite volume massive sine-Gordon model with $N=2$ supersymmetry.
The lattice analog of this continuous quantum field theory
corresponds to a special case of Baxter' famous solvable
eight-vertex lattice model \cite{Bax72}.
In this paper we continue our study \cite{BM05} of this special model
on a finite lattice and unravel its deep connections with Painlev\'e VI theory.

We consider the eight-vertex model on a square lattice with an
odd number, $N=2n+1$, of columns and periodic boundary conditions.
The eigenvalues of the row-to-row transfer matrix of the model, $T(u)$,
satisfy the TQ-equation \cite{Bax72},
\beq
T(u)\,Q(u)=\phi(u-\eta)\,Q(u+2\eta)+\phi(u+\eta)\,Q(u-2\eta),  \label{TQ}
\eeq
where $u$ is the spectral parameter,
\beq
\phi(u)=\vt_1^N(u\,|\,\q), \qquad \q=e^{{\rm i} \pi \tau},
\qquad \rm{Im}\,\tau>0,\label{phi-def}
\eeq
and $\vt_1(u\,|\,\q)$ is the standard theta-function with the periods
$\pi$ and $\pi \tau$ (we follow the notation of \cite{WW}). Here we consider a
special case $\eta=\pi/3$,
where the ground state eigenvalue has  
a very simple form \cite{Bax89,Str01a} for all (odd) $N$
\beq
T(u)=\phi(u), \qquad \eta=\frac{\pi}{3}.\label{T-simple}
\eeq
The equation \eqref{TQ} with this eigenvalue, $T(u)$,
has two different solutions \cite{BLZ97a,KLWZ97}, $Q_\pm(u)\equiv
Q_\pm(u,\q,n)$, which are entire
functions of the variable $u$ and obey the following periodicity
conditions \cite{Bax72,McCoy2}\footnote{The
factor $(-1)^n$ in \eqref{Qper} and
  \eqref{Qper3} reflects our convention for labeling the
  eigenvalues for different $n$}
\beq
Q_\pm(u+\pi)=\pm (-1)^n Q_\pm(u),
\qquad Q_\pm(u+\pi\tau)=\q^{-N/2}\ e^{-iNu}\ Q_\mp(u),\qquad
Q_\pm(-u)=Q_\pm(u). \label{Qper}
\eeq
The above requirements uniquely determine $Q_\pm(u)$ to within a common
$u$-independent normalization factor. It is convenient to
rewrite the functional equation \eqref{TQ} for $Q_\pm(u)$ in the form
\beq
\f_\pm(u)+\f_\pm(u+\frac{2\pi}{3})+\f_\pm(u+\frac{4\pi}{3})=0,\label{TQ1}
\eeq
where
\beq
\f_\pm(u)\equiv\f_\pm(u,\q,n)=
\frac{\vt_1^{2n+1}(u\,|\,\q)}{\vt_1^n(3u\,|\,\q^3)}Q_\pm(u,\q,n),\label{fpm}
\eeq
are the meromorphic functions of the variable $u$ for any fixed
values of $\q$ and $n$. As shown in \cite{BM05} these functions
satisfy the non-stationary Lam\'e equation
\beq
6\,\q\frac{\partial}{\partial\/ \q}\f(u,\q,n)=
\Big\{-\frac{\partial^2}{\partial u^2}+
9\, n\, (n+1)\,\Wp(3u\,|\,\q^3)+c(\q,n)\Big\}\f(u,\q,n),
\label{lame-pde}\eeq
where
the elliptic Weierstrass $\Wp$-function, $\Wp(v\,|\,e^{i\pi\epsilon})$,
has the periods $\pi$ and $\pi\epsilon$ \cite{WW}.
The constant $c(\q,n)$  appearing in \eqref{lame-pde}
is totally controlled by the ($u$-independent) normalization of $Q_\pm(u)$.
Note that this equation (in fact, a more general equation, usually called the
non-autonomous Lam\'e equation)  arose previously
\cite{EK94,KZ84,Ber88,Ols04} in different contexts. We will not explore
these connections here.

Let us quote other relevant results of \cite{BM05}.
Define the combinations
\beq
Q_1(u)=(Q_+(u)+Q_-(u))/2,\quad Q_2(u)=
\,(Q_+(u)-Q_-(u))/2. \label{Q12-def}
\eeq
such that
\beq
Q_{1,2}(u+\pi)=(-1)^n\,Q_{2,1}(u).\label{Qper3}
\eeq
Bearing in mind this simple relation we will only consider
$Q_1(u)$, writing it as $Q^{(n)}_1(u)$ to indicate the
$n$-dependence.  Obviously the linear relations \eqref{TQ1} and
\eqref{lame-pde} remain unaffected if $Q_\pm(u)$ in \eqref{fpm} is
replaced by $Q^{(n)}_1(u)$. The partial differential equation
\eqref{lame-pde} has, of
course, many solutions. Here we are only interested in the very
special solutions, relevant to our original problem of the
eight-vertex model (by definition the functions $Q^{(n)}_1(u)$
are entire quasi-periodic function of $u$ with the periods $\pi$ and
$2\pi\tau$). Introduce new variables $\g$ and $x$, instead of
$\q$ and $u$,
\beq
\g\equiv\g(\q)=
-\left[\frac{\vt_1(\pi/3\,|\,\q^{1/2})}
{\vt_2(\pi/3\,|\,\q^{1/2})}\right]^2,\qquad
x=\g
\left[\frac{\vt_3({u}/{2}\,|\,\q^{1/2})}
{\vt_4({u}/{2}\,|\,\q^{1/2})}\right]^2,\label{gamma}
\eeq
and new functions $\P_n(x,z)$ instead of  $Q^{(n)}_1(u)$,
\beq
Q^{(n)}_1(u)={\mathcal N}(\q,n)\, \vt_3(u/2\,|\,\q^{1/2})\>
\vt_4^{\>2n}(u/2\,|\,\q^{1/2})\>
\P_n(x,z), \qquad z=\g^{-2},\label{Q1}
\eeq
where ${\mathcal N}(\q,n)$ is an arbitrary normalization factor (which
remains at our disposal). The properties of the functions $\P_n(x,z)$
corresponding to the required solutions of \eqref{TQ1} and
\eqref{lame-pde} can be summarized by the following
\begin{conj} {\ }
\begin{enumerate}[(a)]
\item
The functions $\P_n(x,z)$ are polynomials in $x,z$ of the degree $n$ in $x$,
\beq
\P_n(x,z)=\sum_{k=0}^n r^{(n)}_k(z)\,x^k.\label{P-def}
\eeq
while  $r^{(n)}_i(z)$, $i=0,\ldots,n$,
are polynomials in $z$ of the degree
\beq
\deg [r^{(n)}_k(z)]\le \big\lfloor\,{n(n-1)}/{4}+{k}/{2}\,\big\rfloor
\eeq
with {\bf positive integer}\  coefficients.
The normalization of
$\P_n(x,z)$ is  fixed by the requirement \\$r^{(n)}_n(0)=1$ and $\lfloor
x\rfloor$
denotes the largest integer not exceeding $x$.
\item
The coefficients of the lowest and highest powers in
$x$, corresponding to $k=0$ and $k=n$ in \eqref{P-def}, read
\beq
r_0^{(n)}(z)=\tau_n(z,-1/3),\quad r_n^{(n)}(z)=\tau_{n+1}(z,1/6)\ ,
\label{r-coeff}
\eeq
where the functions $\tau_n(z,\xi)$ (for each fixed value of the
their second argument $\xi$) are determined by the recurrence
relation
\bea
&2z(z-1)(9z-1)^2[\log \tau_n(z)]''_z+2(3z-1)^2(9z-1)[\log \tau_n(z)]'_z+\nonumber\\
&
{\ds+
8\Bigl[2n-4\xi-\frac{1}{3}\Bigl]^2{\ds\frac{\tau_{n+1}(z)
\tau_{n-1}(z)}{\tau_n^2(z)}}-}\nonumber&\\
&{-[12(3n-6\xi-1)(n-2\xi)+(9z-1)(n-1)(5n-12\xi)]=0},\label{tau-rec}&
\eea
with the initial condition
\beq
\tau_0(z,\xi)=1,\quad \tau_1(z,\xi)=-4\xi+5/3\ .\label{tau-init}
\eeq
The functions $\tau_n(z,\xi)$
are polynomials in $z$
for all $n=0,1,2,\ldots,\infty$.
\end{enumerate}
\end{conj}
The conjecture has been verified
by an explicit calculation \cite{BM05} of $\P_n(x,z)$ for all $n\le50$.

The
polynomials $\P_n(x,z)$ can be effectively calculated using the
algebraic form (see \cite{BM05}) of the equation (\ref{lame-pde}). The
first few of them read\footnote{The higher
  polynomials with $n\le50$ are available in electronic form upon
  request to the authors.}
\beq
\P_0(x,z)=1, \quad \P_1(x,z)=x+3,\quad
\P_2(x,z)=x^2(1+z)+5x(1+3z)+10,\label{gen4}
\eeq
\beq
\P_3(x,z)=x^3(1+3z+4z^2)+7x^2(1+5z+18z^2)+7x(3+19z+18z^2)+35+21z.\label{gen5}
\eeq
As explained in \cite{BM05}, Eq.\eqref{lame-pde}
leads  to a
descending recurrence relations for the coefficients in \eqref{P-def}, in
the sense that each coefficient $r^{(n)}_k(z)$ with $k<n$ can be
recursively calculated in terms of $r^{(n)}_m(z)$, with $m=k+1,\ldots,n$
and, therefore, can be eventually expressed through the coefficient
$r^{(n)}_n(z)$ of the leading power of $x$. The conditions
that this procedure truncates (and thus defines a polynomial, but not an
infinite series in negative powers of $x$) completely determine the
  starting leading coefficient as a function of $z$.  The
  above conjecture implies that these truncation conditions are
  equivalent to the particular case ($\xi=1/6$) of recurrence
  relation \eqref{tau-rec}, \eqref{tau-init}. A similar statement
  applies to the (smallest power in $x$)
coefficient $r^{(n)}_0$ of $\P_n(x,z)$. The results
  quoted above are due to \cite{BM05} (except for the expression
  for $r^{(n)}_0$ in \eqref{r-coeff} , which is new).

In this paper we show that the recurrence relation \eqref{tau-rec}
exactly coincides with that for the tau-functions associated with
special elliptic solutions of the Painlev\'e VI equation,
revealing hitherto unknown connections of the eight-vertex model and
the non-stationary Lam\'e equation
\eqref{lame-pde} to Painlev\'e VI theory.

Before concluding this introduction, let us
briefly mention some problems related to the special eight-vertex
model considered here. In the trigonometric limit ($\q\to0$)
this model reduces to the special six-vertex model (with the parameter
$\Delta=-1/2$), which is closely
connected with various interesting combinatorial problems
\cite{SR2001,BdGN,Zub04,dFZ05},
particularly, with the theory of alternating sign matrices.
The fact that the polynomials $\P_n(x,z)$ have {\em positive integer
  coefficients}
makes it plausible to suggest that these coefficients have some (yet unknown)
combinatorial interpretation.

In the scaling limit
\beq
n\to \infty, \qquad \q\to 0,\qquad t=8\,n
\q^{3/2}=\textrm{fixed},\label{scaling}
\eeq
the functions \eqref{fpm} essentially reduce to the ground state eigenvalues
$\QQ_\pm(\t)\equiv\QQ_\pm(\t,t)$ of the ${\bf Q}$-operators
\cite{BLZ97a,Fen99} of the
restricted massive sine-Gordon model (at the so-called, super-symmetric point)
on a cylinder of the spatial circumference $R$, where $t=M R$ and $M$ is the
soliton mass and the variable $\t$ is defined as $u=\pi\tau/2-i\t/3$.
With a suitable $t$-dependent normalization of $\QQ_\pm(\t)$,
Eq.\eqref{lame-pde} then reduces to the
``non-stationary Mathieu equation'' \cite{BM05}
\beq
t\frac{\partial}{\partial t}\
\QQ_\pm(\t,t)=\Big\{\frac{\partial^2}{\partial \t^2}-
{\frac{1}{8}}\,t^2\, (\cosh 2\t-1) \Big\}\ \QQ_\pm(\t,t).
\label{mathieu-pde}
\eeq
This equation determines the
asymptotic behavior of $\QQ_\pm(\t)$ at large
$\t$
\beq
\log\, \QQ_\pm(\t)=-\frac{1}{4}\,t\, e^\t+\log\Dcal_\pm(t)
+2\left(\partial_t\log \Dcal_\pm(t)-t/8\right)\,e^{-\t}
+O(e^{-2\t}),\qquad \t\to+\infty,\label{asymp}
\eeq
where $\Dcal_\pm(t)$ are the Fredholm determinants, which previously appeared
in connection with the ``supersymmetric index''
and the dilute polymers on a cylinder \cite{CF92,Zam94,FS92,TW96}.
Note, in particular,
that the quantity
\beq
F(t)=\frac{d}{dt}\,\,U(t), \qquad U(t)=\log\,\frac{\Dcal_+(t)}
{\Dcal_-(t)}\ ,
\eeq
describes the free energy of a single incontractible polymer loop and
satisfies the Painlev\'e III equation \cite{FS92}
\beq
\frac{1}{t}\frac{d}{dt}t\frac{d}{dt}\, U(t)=\frac{1}{2}\,\sinh 2 U(t).
\eeq
\section{Painlev\'e VI equation}

The Painleve ${\bf P}_{ VI}(\a,\b,\g,\d)$  is the following second
order differential equation
\cite{Pain02,Gamb10}
\bea
&{\ds
q''(t)=\frac{1}{2}\biggl(\frac{1}{q(t)}
+\frac{1}{q(t)-1}+\frac{1}{q(t)-t}\biggr)q'(t)^2-
\biggl(\frac{1}{t}+\frac{1}{(t-1)}+\frac{1}{q(t)-t}\biggr)q'(t)+}&\nonumber\\
&{\ds+
\frac{q(t)(q(t)-1)(q(t)-t)}{t^2(t-1)^2}\biggl[\a+\b\frac{t}{q(t)^2}+
\g\frac{t-1}{(q(t)-1)^2}+\d\frac{t(t-1)}{(q(t)-t)^2}\biggr]
}&\ ,\label{pain1}
\eea
where the following parameterizations of four constants \cite{Ok87} is chosen
\beq
\a=\frac{1}{2}\kappa_\infty^2,\quad
\b=-\frac{1}{2}\kappa_0^2,\quad
\g=\frac{1}{2}\kappa_1^2,\quad
\d=\frac{1}{2}(1-\theta^2),\label{pain2}
\eeq
\beq
\kappa_0=b_1+b_2,\quad \kappa_1=b_1-b_2,
\quad \kappa_\infty=b_3-b_4,\quad \theta=b_3+b_4+1.\label{pain3}
\eeq
This equation is equivalent to the Hamiltonian system $H_{VI}(t;q,p)$
described by the equations
\beq
\frac{dq}{dt}=\frac{\partial H}{\partial p},
\quad \frac{dp}{dt}=-\frac{\partial H}{\partial q}\ ,\label{pain4}
\eeq
with
the Hamiltonian function
\bea
&{\ds H_{VI}(t;q,p)=\frac{1}{t(t-1)}\Bigl[q(q-1)(q-t)p^2-}&\nonumber\\
&{\ds -\{\kappa_0\,(q-1)(q-t)+
\kappa_1\,q(q-t)+(\theta-1)q(q-1)\}p+\kappa\>(q-t)],}&\label{pain5}
\eea
where $q\equiv q(t)$, $p\equiv p(t)$ and
\beq
\kappa=\frac{1}{4}(\kappa_0+\kappa_1+\theta-1)^2
-\frac{1}{4}\kappa_\infty^2.\label{pain6}
\eeq
One can introduce an auxiliary Hamiltonian $h(t)$,
\beq
h(t)=t(t-1)H(t)+e_2(b_1,b_3,b_4)
\,t-\frac{1}{2}e_2(b_1,b_2,b_3,b_4)\label{pain7}
\eeq
where $e_i(x_1,\ldots,x_n)$ is the $i$-th elementary symmetric
function in $n$ variables.

Okamoto \cite{Ok87} showed that for each pair $\{q(t),p(t)\}$
satisfying (\ref{pain4}), the function $h(t)$ solves
the ${\bf E}_{{VI}}$ equation,
\beq
h'(t)\Bigl[t(1-t)h''(t)\Bigr]^2
+\Bigl[h'(t)[2h(t)-(2t-1)h'(t)]+b_1b_2b_3b_4\Bigr]^2=
\prod_{k=1}^4\Bigl(h'(t)+b_k^2\Bigr)\ ,\label{pain8}
\eeq
and $q(t)$ solves ${\bf P}_{ VI}(\a,\b,\g,\d)$,  (\ref{pain1}).

Conversely, for each solution $h(t)$ of (\ref{pain8}),
such that $\frac{d^2}{dt^2}h(t)\neq0$, there exists
a solution $\{q(t),p(t)\}$ of (\ref{pain4}), where $q(t)$ solves (\ref{pain1}).
An explicit correspondence between three sets $\{q(t),q'(t)\}$,
$\{q(t),p(t)\}$ and $\{h(t),h'(t),h''(t)\}$
is given by birational transformations,
which can be found in \cite{Ok87}.

The equation ${\bf E}_{{VI}}$  is
a particular case of a more general class of the second
order second degree equation
\beq
y''(t)^2=F(t,y(t),y'(t))\label{pain9}
\eeq
where $F$ is rational in $y(t)$, $y'(t)$,
locally analytic in $t$, with the property that the only movable
singularities of $y(t)$ are poles.
Such equations were classified in \cite{Cos93}, where Eq.\eqref{pain8}
is refered to the ``SD-I type''.
Originally, we have obtained equations of the form (\ref{pain9})
for asymptotics of the polynomals $\P_n(x,z)$ and
only then reduced them to (\ref{pain8}) and (\ref{pain1}).

The group of Backlund transformations of ${\bf P}_{VI}$ is
isomorphic to the affine Weyl group of the type $D_4$. It contains
the following transformations of parameters (only 5 of them are independent)
\beq
w_1: b_1\leftrightarrow b_2,\quad w_2:
b_2\leftrightarrow b_3,\quad w_3: b_3\leftrightarrow b_4,
\quad w_4: b_3\to -b_3,\> b_4\to-b_4,\label{pain10}
\eeq

\beq
x^1: \kappa_0\leftrightarrow\kappa_1,\quad
x^2: \kappa_0\leftrightarrow\kappa_\infty,\quad
x^3: \kappa_0\leftrightarrow\theta\label{pain11}
\eeq
and the parallel transformation
\beq
l_3: {\bf b}\equiv(b_1,b_2,b_3,b_4)
\to{\bf b^+}\equiv(b_1,b_2,b_3+1,b_4).\label{pain12}
\eeq
Here we shall use only two transformations: $x^2$ and $l_3$.
The canonical transformation $x^2$ corresponds to a simple change of
variables \cite{Ok87},
\beq
x^2_\star: (q,p,H,t)\to(q^{-1},\epsilon q-q^2 p,-H/t^2,1/t),\quad
\epsilon=\frac{1}{2}
(\kappa_0+\kappa_1+\theta-1+\kappa_\infty)\ .\label{pain13}
\eeq
The birational canonical transformation,
\beq
\{q,p\}=\{q({\bf b}),p({\bf b})\}
\to\{q^+,p^+\}=\{q({\bf b^+}),p({\bf b^+})\}\ ,\label{pain14}
\eeq
corresponding to $l_3$, can be obtained from the observation that \cite{Ok87}
\beq
h^+(t,q^+,p^+)=h(t,q,p)-q(q-1)p+(b_1+b_4)q
-\frac{1}{2}(b_1+b_2+b_4).\label{pain15}
\eeq
Here we shall only give the expression for $q$ in terms of $\{q^+,p^+\}$
\bea
 q\Bigl\{p_+^2q_+(q_+-1)(q_+-t)+
+p_+\bigl[2(1+b_1+b_3)q_+
(1-q_+)+(b_1+b_2)(q_+-t)+2b_1(t-1)q_+\bigr]+&&\nonumber\\
+(1+b_1+b_3)\bigl[(1+b_1+b_3)(q_+-1)+t(1+b_3-b_1)+b_1-b_2\bigr]\Bigr\}
-&&\nonumber\\
-t\bigl[p_+(q_+-1)-1-b_1-b_3\bigr]\bigl[p_+q_+(q_+-1)
-(1+b_1+b_3)q_++b_1+b_2\bigr]=0.&&\label{pain16}
\eea
Later we will need the transformation $x^2 l_3 x^2$
\beq
x^2 l_3 x^2: \{{\bf b}\}=\{b_1,b_2,b_3,b_4\}\to
\{{\bf\tilde b}\}=\{b_1+\frac{1}{2},b_2
+\frac{1}{2},b_3+\frac{1}{2},b_4+\frac{1}{2}\}.
\label{pain16a}
\eeq
Combining (\ref{pain13}-\ref{pain16}) one obtains
\bea
{q=
\bigl\{\tilde p^2(\tilde q-t)(\tilde q-1)+
\tilde p\,(2+b_1+b_2+b_3+b_4+(b_1-b_2)t
-(2+2b_1+b_3+b_4)\tilde q)}\nonumber\\
{+(1+b_1+b_3)(1+b_1+b_4)\bigr\}/
\bigl\{(1+b_1+b_3+\tilde p\,(1-\tilde q))(1+b_1+b_4+\tilde p
\,(1-\tilde q))\bigr\}}\ ,
\eea
where $q=q({\bf b})$ and  $\tilde q=q({\bf\tilde b})$, $\tilde
p=p({\bf\tilde b})$.

\section{Special elliptic solutions}

It goes back to Picard \cite{Pic89} that if parameters \eqref{pain3}
satisfy
\beq
b_1=b_2=0,\quad b_3=b_4=-1/2,\label{pain17}
\eeq
then a general solution of ${\bf P}_{VI}$ reads (see also \cite{Maz01})
\beq
q(t)=\textrm{\Large$\wp$}(c_1\omega_1+c_2\omega_2;\omega_1,\omega_2)
+\frac{t+1}{3}\label{pain18}
\eeq
where {\Large$\wp$}$(u;\omega_1,\omega_2)$ is the Weierstrass elliptic function
with half-periods $\omega_{1,2}$, $c_{1,2}$ are arbitrary constants
 and $\omega_{1,2}(t)$ are two linearly independent solutions
of the hypergeometric equations
\beq
t(1-t)\omega''(t)+(1-2t)\omega'(t)-\frac{1}{4}\omega(t)=0.\label{pain19}
\eeq
It is convenient to choose
\beq
\omega_1(t)=\frac{\pi}{2}\phantom{|}_2F_1(\frac{1}{2},\frac{1}{2};1;t)=\mbox{\bf K}(t^{1/2}),\quad
\omega_2(t)=i\frac{\pi}{2}\phantom{|}_2F_1(\frac{1}{2},\frac{1}{2};1;1-t)=i\mbox{\bf K}'(t^{1/2}),
\label{pain20}
\eeq
where
{\bf K} and $\mbox{\bf K}'$ are the complete elliptic integrals of the modulus $k=t^{1/2}$.

Using expressions for invariants of the Weierstrass function $e_1, e_2, e_3$
\beq
e_1=1-\frac{t+1}{3},
\quad e_2=t-\frac{t+1}{3},\quad e_3=-\frac{t+1}{3}\label{pain21}
\eeq
we can rewrite Picard's solution of ${\bf P}_{VI}$ as
\beq
q(t)=\mbox{ns}^2(c_1\mbox{\bf K}+c_2i\mbox{\bf K}',k),
\quad k=t^{1/2} \ .\label{pain22}
\eeq
Let us choose
\beq
c_1=1,\quad c_2=\frac{1}{3}\label{pain23}
\eeq
and denote corresponding solution of (\ref{pain1}) as $q_1(t)$.
Using addition theorems for elliptic functions it is easy to show that
$q_1(t)$ satisfies the algebraic equation
\beq
q_1^4(t)-4t\,q_1^3(t)+6t\,q_1^2(t)-4t\,q_1(t)+t^2=0.\label{pain24}
\eeq
In fact, this algebraic solution of ${\bf P}_{VI}$ is very
special. It solves (\ref{pain1}) not only for the Picard's choice of
parameters (\ref{pain17}) but also for
\beq
b_1=\xi-\frac{1}{6},\quad b_2=0,\quad b_3=\xi-\frac{2}{3},
\quad b_4=2\xi-\frac{5}{6},\label{pain25}
\eeq
where $\xi$ is an arbitrary parameter. This happens because
${\bf P}_{VI}$, (\ref{pain1}), splits up into two different equations which
are both satisfied by (\ref{pain24}).

Using above formulas we can easily find expressions for $p_1(t)$ and
$h_1(t)$ corresponding (\ref{pain24})
\beq
 p_1(t)=\frac{(1-3\xi)(3t-2tq_1(t)-q_1^2(t))}{6t(q_1(t)-1)^2},\quad
h_1(t)=\frac{1-2t}{72}+\frac{(1-2\xi)(1-3\xi)}{4}
\frac{(t+q_1(t)-2t\,q_1(t))}{q_1(t)-t}.\label{pain27}
\eeq
Now we shall assume that this solution corresponds to the case $n=1$
and apply Backlund transformation $x^2\, (l_3)^{1-n}x^2$ to obtain a
series of solutions $\{q_n(t),p_n(t),h_n(t)\}$ with parameters
\beq
b_1=\frac{1}{3}+\xi-\frac{n}{2},\quad b_2=\frac{1}{2}-\frac{n}{2},\quad
b_3=-\frac{1}{6}+\xi-\frac{n}{2},\quad b_4=
-\frac{1}{3}+2\xi-\frac{n}{2}.\label{pain28}
\eeq
At this stage we are ready to establish a connection with the
elliptic parametrization from the first section.
Let us assume that the elliptic nome $\q$ in
(\ref{phi-def}), (\ref{fpm}-\ref{lame-pde}) and (\ref{gamma})
is given by
\beq
\q=\exp\Bigl\{i\pi\frac{2}{3}\frac{\mbox{\bf K}'(k)}
{\mbox{\bf K}(k)}\Bigr\},\quad k=t^{1/2}, \label{nome}
\eeq
where $\mbox{\bf K}(k)$ and $\mbox{\bf K}'(k)$ are defined by (\ref{pain20}).

Using Landen transformation for elliptic functions
it is easy to obtain the following rational
parametrization for $z=1/\g(\q)^2$ defined in (\ref{gamma}), (\ref{Q1}),
and for $t$, $q_1(t)$ in (\ref{pain24}) 
\beq
z=\frac{1}{\g^{2}(\q)}=\frac{1+\ov\g}{(3-\ov\g)\ov\g},\quad
t=\frac{(1-\ov\g)(3+\ov\g)^3}{(1+\ov\g)(3-\ov\g)^3}
,\quad q_1(t)=\frac{(1-\ov\g)(3+\ov\g)}{(1+\ov\g)(3-\ov\g)},
\label{pain30}
\eeq
in terms of a new parameter
\beq
\ov\g\equiv\g(\q^{1/2})\ ,\label{sgamma}
\eeq
defined by \eqref{gamma} with $\q$ replaced by $\q^{1/2}$.
Note that such parameterization of Picard's solutions 
of ${\bf P}_{VI}$ with the above choice \eqref{pain23}
of the parameters $c_1$ and $c_2$ has already
appeared in \cite{Dub96,Maz01}.

{}From these formulas one can get an explicit connection of variables
$t$ and $z$
\beq
t=\frac{(z-1)(1-9z)^3}{32z}\biggl[1+\frac{27z^2-18z-1}
{\sqrt{(1-z)(1-9z)^3}}\biggr].\label{pain31}
\eeq

Now we can construct a sequence of $\tau$-functions associated with a
series of auxiliary Hamiltonians $h_n(t)$. It appears that
corresponding $\tau$-functions are polynomials in variable $z$.

First, let us introduce a sequence of functions $\sigma_n(z)$
considering them as functions of $z$
\bea
\ds{\sigma_n(z)=\frac{1}{tz}\sqrt{\frac{1-9z}{1-z}}\biggl\{h_n(t)
+\frac{1}{72}(2t-1)+
(n-1)^2\Bigl[\frac{t-1}{4}+\frac{1-9z}{8}\sqrt{(1-t)z}\>\Bigr]+
}\label{pain32}\\
\ds{+
(n-1)\bigl(\xi-\frac{5}{12}\bigr)}\Bigl[1-t+\frac{t(1-3z)}
{\sqrt{(1-z)(1-9z)}}\Bigr]
+(\xi-\frac{1}{2})(\xi-\frac{1}{3})
\biggl[\frac{3}{2}-\sqrt{\frac{1-t}{z}}\>\biggr]\biggr\}.\nonumber
\eea
Comparing it with (\ref{pain27}) and using (\ref{pain30}) it is easy
to see that
\beq \sigma_1(z)=0\label{pain33}.
\eeq
Then using Backlund transformations $x^2l_3^{-1}x^2$ and $x^2l_3x^2$
it is not difficult to show that
\beq
\sigma_i(z)=0,\quad i=0,1,2.\label{pain34}
\eeq
The easiest way to do that is to calculate $h_i(t)$ , $i=0,2$ in terms
of $h_1(t)$ (\ref{pain27}), substitute into (\ref{pain32}) and use a
rational parametrization (\ref{pain30}).

Now let us introduce a family of $\tau$-functions $\tau_n(z,\xi)$ via
\beq
\sigma_n(z)=\frac{d}{dz}[\log \tau_n(z,\xi)]\label{pain35}
\eeq
and fix a normalization for $n=0,1,2$ as
\beq
\tau_0(z,\xi)=1,\quad \tau_1(z,\xi)=-4\xi+5/3,\quad
\tau_2(z,\xi)=3(2\xi-1)(3\xi-1).\label{pain36}
\eeq
Using Okamoto's Toda-recursion relations for $\tau$-functions
for ${\bf P}_{VI}$,  generated via successive applications of parallel
transformation $l_3$ \cite{Ok87}, one can show the recurrence
relation for $\tau_n(z,\xi)$ exactly coincides with \eqref{tau-rec}.
Thus, we showed that the leading coefficient
and the constant term  of $\P_n(x,z)$ (considered as polynomials in
$x$) can be expressed in terms $\tau$-functions for special solutions
of ${\bf P}_{VI}$.

At the moment we do not have a complete proof of the polynomiality of
$\tau_n(z)$. Note, that this property takes place
provided that two successive $\tau$-functions $\tau_n(z)$ and
$\tau_{n+1}(z)$ do not have a nontrivial common divisor (which is a
polynomial in $z$).

One of the challenging problems is to find a determinant
representation for $\tau_n(z,\xi)$
similar to that known other polynomial solutions of the Painlev\'e equations.
It could help to clarify the structure of  $\P_n(x,z)$ and possibly to
establish a connection with problems of combinatorics.

\section*{Acknowledgments}

The authors thank M.T.~Batchelor, P.J.~Forrester, N.~Frankel,
I.M.~Krichever, S.L.~Lukyanov, S.M.~Sergeev,
Yu.G.~Stroganov, N.S.~Witte, A.B.~Zamolodchikov, Al.B.~Zamolodchikov and
Z.-B.~Zuber for
stimulating discussions and valuable remarks.

\end{document}